# Compressive Sensing for Dynamic XRF Scanning

George Kourousias[a]*, Fulvio Billè[a], Roberto Borghes[a], Antonio Alborini[b], Simone Sala[a,c], Roberto Alberti[b] and Alessandra Gianoncelli[a]

[a] Elettra Sincrotrone Trieste, Basovizza, Trieste, Italy
[b] XGLab srl, Bruker, Bruker Nano Analytics, Milan, Italy
[c] now at MAXIV, Lund, Sweden

Correspondence email: george.kourousias@elettra.eu

**Synopsis**: *Sparse and dynamic XRF scanning enables much larger and faster acquisitions. Through suitable Compressive Sensing, certain prohibitively difficult experiments become feasible.*

**Abstract (max 200 words)**: X-Ray Fluorescence (XRF) scanning is a widespread technique of high importance and impact since it provides chemical composition maps crucial for several scientific investigations. There are continuous requirements for larger, faster and highly resolved acquisitions in order to study complex structures. Among the scientific applications that benefit from it, some of them, such as wide scale brain imaging, are prohibitively difficult due to time constraints. However, typically the overall XRF imaging performance is improving through technological progress on XRF detectors and X-ray sources. This paper suggests an additional approach where XRF scanning is performed in a sparse way by skipping specific points or by varying dynamically acquisition time or other scan settings in a conditional manner. This paves the way for Compressive Sensing in XRF scans where data are acquired in a reduced manner allowing for challenging experiments, currently not feasible with the traditional scanning strategies. A series of different compressive sensing strategies for dynamic scans are presented here. A proof of principle experiment was performed at the TwinMic beamline of Elettra synchrotron. The outcome demonstrates the potential of Compressive Sensing for dynamic scans, suggesting its use in challenging scientific experiments while proposing a technical solution for beamline acquisition software.

## Introduction

X-ray Fluorescence (XRF) spectroscopy and imaging are well established techniques used in a wide range of scientific applications[1–3]. In the case of imaging, the XRF emission signal is acquired for each position of a scanned sample, resulting, after suitable analysis, into a set of elemental maps. This combination of elemental information with spatial distribution provides important information on the analysed areas but being based on a scanning acquisition mode implies that large maps (>1M points) may require long acquisition times. In the soft X-ray regime also known as Low Energy XRF (LEXRF) the acquisition times are even longer due to the low fluorescence yield below 2 keV. LEXRF on the other hand is crucial for studies and applications that probe light elements, such as C, N, O, Na, Mg, Al, Si and P, important in biological systems, and transition metals (Mn, Fe, Ni, Co, Cu and Zn). Such applications are often biomedical and pose challenges when it comes to imaging large area samples like brain tissues[4–6] or monitoring fast dynamic phenomena like electrochemical processes[7–9]. Improvements in LEXRF detectors and soft X-ray sources together with advanced instrumentation (i.e. better vacuum conditions, larger detection solid angle, and better motor stages) pave the way for such large scans that may enable new science. Still additional approaches may be needed as the aforementioned improvements may not translate directly into reduced measurement times. Table 1 shows typical acquisition time duration and scan range in pixels for the LEXRF system installed at the TwinMic beamline[10] (case a). It also shows that by increasing the number of pixels by 100 times (from 100x100 to 1000x1000), even by reducing the acquisition time by 3 times (case b), the total duration of the scan can reach forbiddingly high numbers compared to the duration of a beamtime experiment. However if the acquisition is based on sparse and compressive sensing approaches such those this paper is proposing (case c), the total measurement can be rendered feasible. Thus, this manuscript suggests reduced and selective ways for acquiring LEXFR maps, which can be obviously applied to any energy range. It is based on Compressive Sensing[11,12] which is an emerging very effective technique for reconstruction from a relatively small number of data samples without compromising the imaging quality; indeed in our case it allows performing scans of a dynamic nature where it is possible to skip points (sparse) and acquire with variable parameters (i.e. acquisition time). Such an approach may lead to a substantial reduction of the required time. This reduction in terms of time, combined with other technological and scientific advances like new detectors, can produce maps of very large areas that could not be acquired with the traditional ways. This paper reports the results of the suggested method applied in a series of LEXRF acquisitions at the TwinMic synchrotron spectromicroscopy beamline[10] of Elettra Sincrotrone Trieste (Trieste, Italy). A flexible and modular beamline control system was used[13] in order to perform Compressive Sensing scans while post-processing software has been developed for the reconstruction of the sparse maps into traditional ones that can be processed with standard analysis software. The used detector system is based on a 8 SDD setup[14,15] with a novel multi-channel analyser[16] (DANTE digital pulse processor, XGLab Bruker).

|   | Width x Height (total scan positions) | Acquisition time (dwell per point) | Total scan time (excluding overhead) | Feasibility |
|---|---|---|---|---|
| a | 100x100 (typical scan) | 3 sec | 8.3 hours | normal |
| b | 1000x1000 (megapixel range) | 1 sec (with new SDDs) | 11.5 days | impractical |
| c | 1000x1000 (proposed method) | Variable 0.5-1sec in sparse scanning at 15% | 10 to 20 hours | feasible |

Table 1. The time required at TwinMic for a typical LEXRF scan of 100x100 positions and 3 sec acquisition per pixel (a) is increasing prohibitively for scans of larger sizes (b). The proposed compressive sensing method can render such scans feasible (c).

**Results and Discussion**
**1. Sparse scans and Masking**

Typical scanning assumes the acquisition of XRF spectra at positions belonging to a regular cartesian grid. This simplifies various processes but most notably allows for the rectangular representation of the data in the form of images or volumes. The first type of reduced acquisition we implemented is Sparse scanning coupled with masking. It consists in two steps: i) manually selecting certain features of the sample that need to be scanned (Masking) at a certain spatial resolution and ii) acquiring data at sample stage steps larger than the focal spot size of the X-ray beam in the remaining masked areas (Sparse Scanning) while at high resolution in the regions of interest. To note, the density of the Sparse scan can also be variable. The workflow of the proposed method is shown in detail in Figures S1 in Supplementary Data through a suitable flowchart scheme. The selection (or masking) is performed through a simple interface (e.g. touch screen, pen stylus, smartphone) on preview data such as a faster scan or transmission (STXM) map. In order to assemble the sparse and masked data so that they can be processed, fitted and visualised by software (i.e. PyMca[17]) in a suitable elemental map, we reconstruct it to a dense matrix by using a suitable in-painting technique[18,19]. It should be noted the PyMca[17] allows for easy visualisation of sparse data without the need to densify the map (Mask Scatter View); still in-painting techniques may improve the visualisation at processing cost. The in-painting fills the missing values of the Sparse Scan by suitable interpolation based on the neighbouring XRF data. The result is a 3-dimensional array (width, height, energy) compatible with the standard processing workflows. A visual example of the described method is presented in figure 1, where a standard XRF map (Figure 1d), in this case of Silicon, acquired point by point in a regular square scan on a foraminifera shell is compared with the same scan acquired with the proposed approach (Figure 1e), based on a mask (Figure 1b) defined from a previously acquired absorption image (Figure 1a), and finally reconstructed (in-painted; Figure 1f) to resemble a full one. The mask is also extended by a set of sparse scan points (Figure 1c), in order to sample the areas outside the regions of interest. The reported example demonstrates the advantage of the method. The less interesting areas (in this case the ultralene foil, that is the sample support and the paraffin embedding medium) are under-sampled while the regions of interest (the shell sections) are scanned with a suitable spatial resolution, allowing to provide the same scientific information with a reduced acquisition time. For this particular system the measurements that lead to Figure 1d required 6 hours, while the ones that produced Figure 1f just 2 hours.

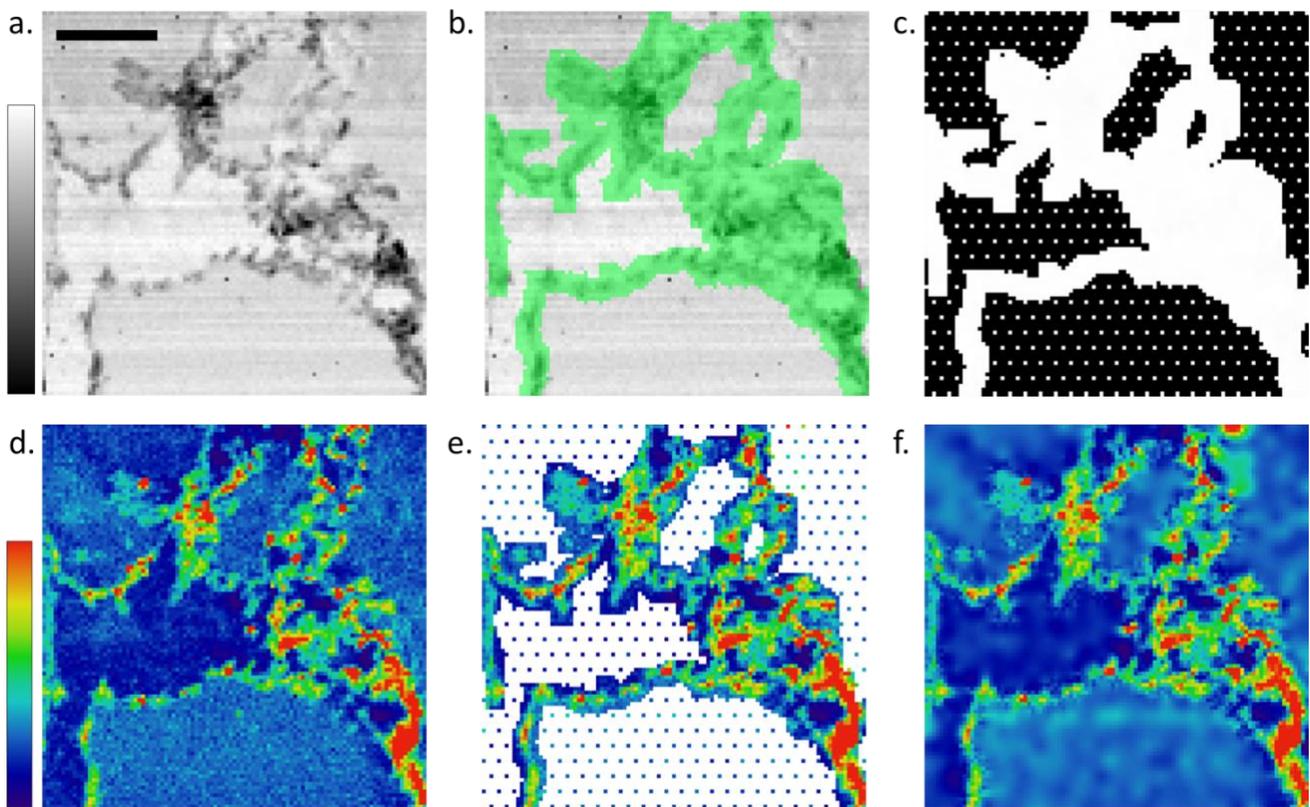

**Figure 1.** A rapidly acquired STXM map (a) (80μm x 80μm, 50x50 pixels, 20ms dwell time/pixel, scale bar=20μm) is used to create a mask (b) which is dense only in the regions of interest (green areas in c). A sub-sampled sparse XRF acquisition (e) is approximately 3 times faster than a full one (d). When the sparse is reconstructed through biharmonic in-painting (f) it can be processed with the usual XRF workflows producing similar results. In this specific case Panels d, e and f shows Si XRF signal collected at 1.95 keV on a foraminifera section over an area of 80μm x 80μm, with 1.6 μm step size and 3s acquisition time/pixel.

## 2. Conditional scans and multimodal acquisitions

Some XRF instruments are embedded in more complex systems, using additional detectors which allow them to operate in a multimodal way. The TwinMic beamline endstation for instance allows for the concurrent acquisition of XRF and STXM data[20] (Figure 2a). In a Conditional Scanning mode, the system decides on-the-fly whether it should acquire XRF data for a given point according to the transmission STXM signal of that point. Note that in many setups, STXM data may be collected 2 orders of magnitude faster than XRF. Such a conditional scanning represents a case of Compressive Sensing by using a fast probing microscopy technique of less costly nature, in terms of time, to decide whether to acquire data in that specific point. The result is processed in a similar manner to the Masked scans where the missing values are interpolated in order to reconstruct a complete matrix.

Figure S1b in Supplementary Data illustrates in detail the proposed acquisition method by a suitable flowchart while Figure 2 depict a successful example of it. Based on the simultaneous acquisition of an absorption image (2b) and on the choice of an absorption threshold, which discriminates the sample from the support and the embedding medium (2c), the XRF signal, of Silicon in this case, is acquired only on specific areas (2d), allowing to speed up the acquisition compared to the standard full area (2e). For this specific example, where soybean root section[21] was imaged, the proposed method allowed to reduce the measurement time by around 66%. In order to obtain a complete image, an in-painting reconstruction is performed *a posteriori* (Figure 2f), as described in the previous section.

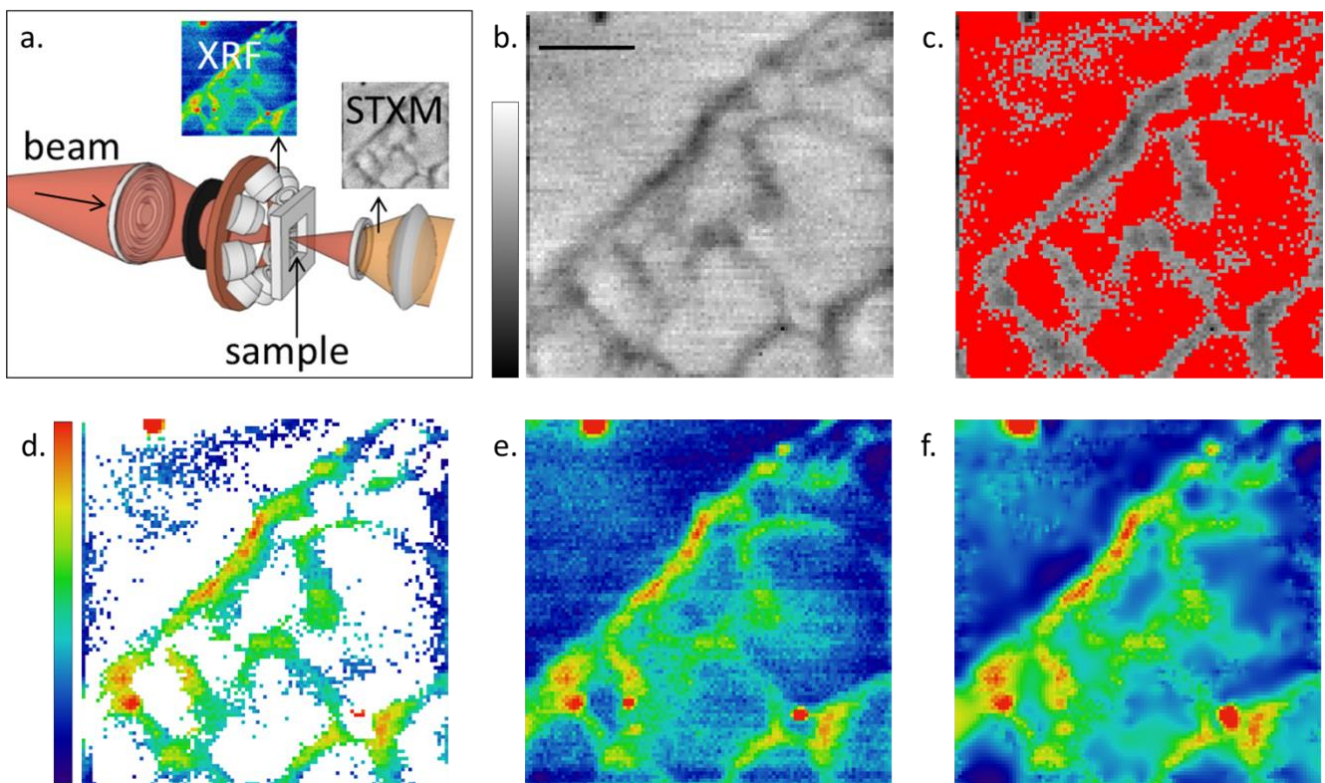

**Figure 2.** A dual XRF+STXM setup (a) allows for using the fast STXM modality (b) (80μm x 80μm, 50x50 pixels, 20ms dwell time/pixel, scale bar=20μm) as a decision factor for performing a slower XRF acquisition (d). The scan positions in red (c) are above a transmission threshold resulting to a fast 66% sub-sampled XRF acquisition (d). The in-painting reconstruction (f) of the sparse map (d) is comparable to a slow full acquisition (e). In this specific case Panels c, d and f shows Si XRF signal collected at 1.95 keV on a foraminifera section over an area of 80μm x 80μm, with 1.6 μm step size and 3s acquisition time/pixel.

## 3. Dynamic XRF scanning

Traditional XRF scans generate maps that are processed as a whole. Step size, spot size, and acquisition time are common parameters across the whole map: they are usually set at the beginning of the acquisition and kept constant for the whole measurement. The proposed Dynamic XRF strategy allows for any parameter to be variable in each position of the scan. The variation (e.g. of acquisition time) at a given scan position depends on the actual XRF signal that is collected there in real-time. During the experiment this compressive sensing approach allows to decrease the acquisition time in the areas of the map which have for instance the composition of the background (i.e. ultralene or $Si_3N_4$) and increases it for better statistics for regions containing the elements of interest. The resulting map requires a normalisation of the acquisition time of each pixel prior to fitting it with the traditional software. The workflow of the proposed method is presented in detail in Figure S1c in Supplementary Data through a general description of the procedures involved. The objective is to increase the acquisition time for the scan positions where a predefined element of interest is present.

Figure 3 showcases an example of the proposed approach on a section of a soybean root specimen[21] acquired with submicrometric spatial resolution. Figure 3a depicts the absorption image of the analysed area, to highlight its morphology, even though it is not used at any stage of the proposed method. The dynamic scan acquisition assumes the definition of an XRF signal threshold for a specific element. This can be retrieved by basic signal integration of more sophisticated fitting. For each position in the scanned area, a fast XRF acquisition is used to decide dynamically whether to acquire a longer XRF signal for better statistics on that position. For this specific case a ROI window was selected across Na peak in the XRF spectrum and the threshold was applied on this signal. The fast acquisition was set to 1s while the longer one to 7s. Where the threshold was met the acquisition was set to 7s instead of 1s. By doing so a set of XRF data were collected at variable exposure times (1 and 7 s) that after a suitable normalisation were reconstructed to a regular map (figure3b). In order to estimate the reduction in time compared to acquiring a full 7s/pixel map, Figure 3c shows in black the pixels which were acquired at 1s and in green the ones at 7 s. Considering that 32% of the areas was acquired at 7 s while the remaining one at 1s, the dynamic scan required 43% of the total time (full XRF map at 7s). For comparing and validating the results, full Na maps were acquired at 1s (Figure 3d) and 7s (Figure 3e). The advantage of the proposed method is evident by comparing Figures 3b and 3e, where the Na signal is the same on the cell walls of the root, i.e. the regions of interest, while noisy (same as in Figure 3d) in the remaining ones. This means that maps with similar information on Na can be obtained by saving 53% of the total time for this given sample.

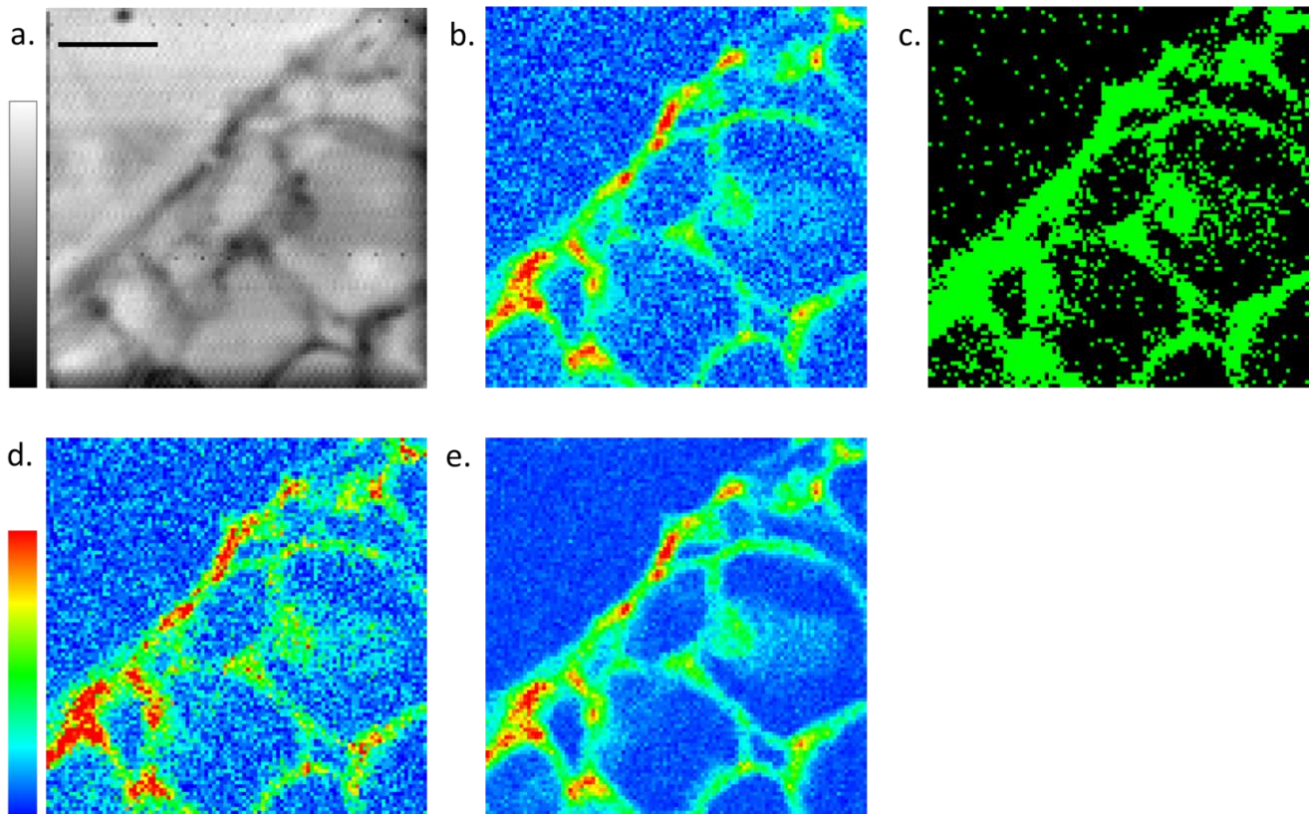

**Figure 3.** (a) STXM absorption image of a 7μm thick soybean root section where the XRF maps were collected (80μm x 80μm, 100x100 pixels, 40ms dwell time/pixel, scale bar=20μm). (b). Resulting dynamic XRF scan where the point indicated in black in panel (c) where acquired with 1s acquisition time, while the one in white with 7s, all normalised to 1s acquisition time. (c) According to the pre-set XRF threshold the points in black were acquired at 1s and the ones in green at 7s. Full XRF map acquired with 1s/pixel (d) and 7s/pixel (e). In this specific case Panels b, d, e show Na XRF signal collected at 1.95 keV over an area of 80μm x 80μm, with 800 nm step size.

The proposed Dynamic XRF strategy could also by extended by considering more than one chemical element of interest, as proposed in Figure 4, where a simplified region scanned in 5x5 steps in shown in panel 1. The aim is to increase the acquisition time for the scan positions where objects characterized by predefined chemical elements are present. The grey areas represent the background substrate while the blue circle and the red rectangular region are the features of interest, with the red region being a higher priority one. This approach assumes that the regions of interest in the specimen are distinguished by at least one specific characteristic element which is not present in the background or in less interesting regions. Panel 2 depicts possible spectra of the three areas, grey, red and blue, plotted with the same colors as the corresponding areas. The element indicated with the red ROI (c) is not present in the grey and blue areas therefore it can be used as fingerprint for the red region in panel 1. This simplified example of a dynamic scan assumes the that higher XRF statistics are required for the red element regions without knowing *a priori* their location in the scan. With this strategy the acquisition time can be increased for instance to 1.2 seconds when the red "peak" (panel 2) is present while the grey areas of the background are fast scanned at 0.2s. Since the blue region is still an area of interest but of less importance than the red region, one can decide to increase the acquisition time for instance to 0.6s when the red peak is absent but the blue one is present. This simple example can be easily extended to more complex methods where ratios of XRF signals are taken into account. Such strategies can be especially useful in XRF scans where there is a hypothesis for the existence of a trace element in an unknown or limited region of the sample.

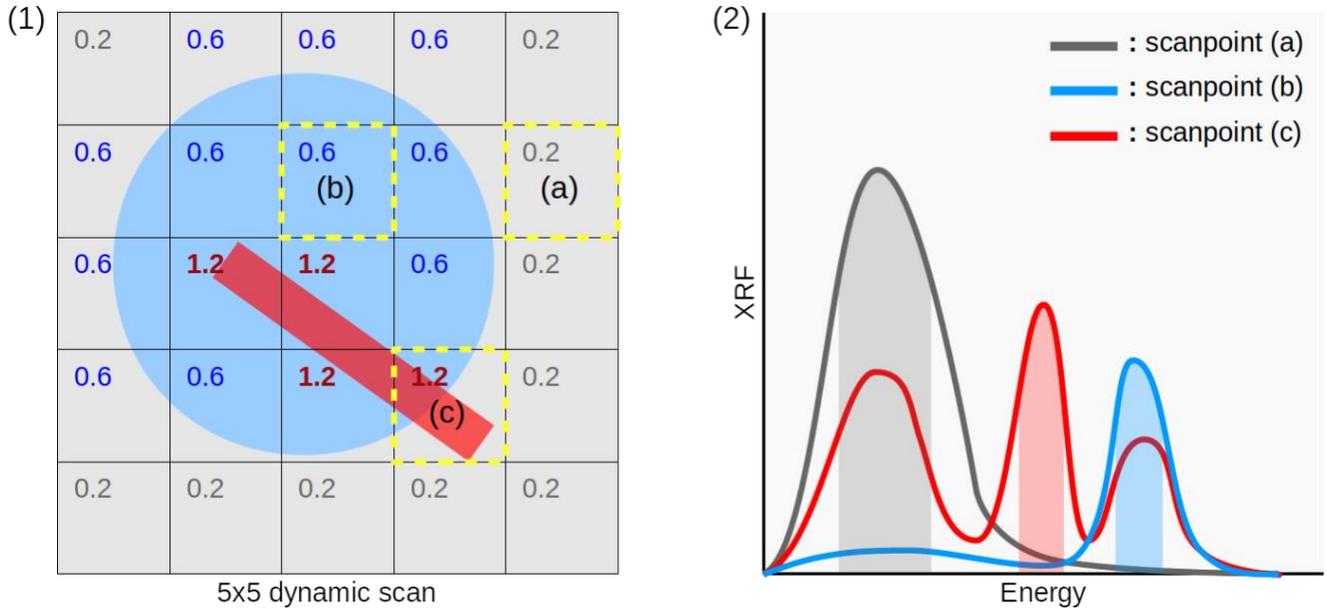

**Figure 4.** Panel (1): a 5x5 pixel XRF scan of a sample containing three characteristic distinct elements (grey, blue, red). The dynamic modality is set to expose longer for the Red element (1.2s) and less for the Blue (0.6s) and Grey (0.2s). Panel (2): XRF spectra of three scanpoints (panel 1: a,b,c in yellow). The presence of Red and/or blue XRF peak in (c) allows for varying dynamically on-the-fly a suitable acquisition time.

## 4. Current Perspective and Future Directions

The proposed methods [sec. 1,2,3] for Compressive Sensing in XRF scans, other than rendering new experiments feasible (large areas, faster scans, more samples), may pave the way for additional advances both scientific and technological. For instance the authors and their associated labs are outlining a pump-probe inspired experiment where the XRF acquisition is subsequent to an excitation (i.e. sample heating through radiation); since these kinds of experiments have specific timing to track transition or changes in the samples, they could really benefit from the proposed approaches. Fast XRF mapping of large sample areas at high spatial resolution could provide multi-scale insights, since it could be applied to different length-scales (from mm to nm) and promote high impact research in multiple scientific fields from material science to medicine, archaeometry and food science. Elemental maps of large areas (mm) of such samples and nanometric resolution could be acquired faster with the dynamic scanning strategies proposed in this manuscript. Moreover, the proposed methods can be applied not only singularly but also as a combination of different ones and they could benefit from Machine Learning techniques. For instance Machine Learning can be used both for in-painting methods (filling in the missing XRF points in sparse scans) and for altering the motor steps of the scan (position jumping). Figure 5 reports an example of a successful combination of methods 1 and 2 on a cane root section. A preliminary mask was selected to cover the root section borders adopting an in-painting strategy for the remaining areas. The mask was deliberately left wider than the actual root borders to compensate for possible sample drifting during the scan. In the non-masked areas methods 2 was applied in order to acquire the XRF signal only for specific absorption signals, in this case the point corresponding to the root. The deployment of methods 2 in the non-masked region assured a faster and safer (free from possible sample drifting) acquisition, then deploying only method 1 alone. Figure 5 depicts the overlap of STXM and LEXRF scans over an area of a 2mm x 0.8mm of the cane root section. The scan resolution using the techniques of this paper varied from 20 to 2 microns on the masked and the finest resolved areas, respectively. This allowed for examination of C and Mg on the border of the root by performing a scan 3 times faster than that of a traditional dense one.

Eventually the proposed techniques can be used for fly scans (the currently used DANTE electronics permits it[16]) and may be extended to variable velocity scans. The increase of speed could be also useful for XRF tomography experiments that tend to be very challenging and time-consuming. An extension of conditional scans and multimodal acquisitions [sec.2] is the use of phase contrast information[22,23] rather than simple STXM transmission as a decisive factor for the XRF acquisition. Finally XRF topography artifacts are a topic of research[24–27] and is only natural to use the phenomenon in the context of Dynamic XRF scanning [sec.3].

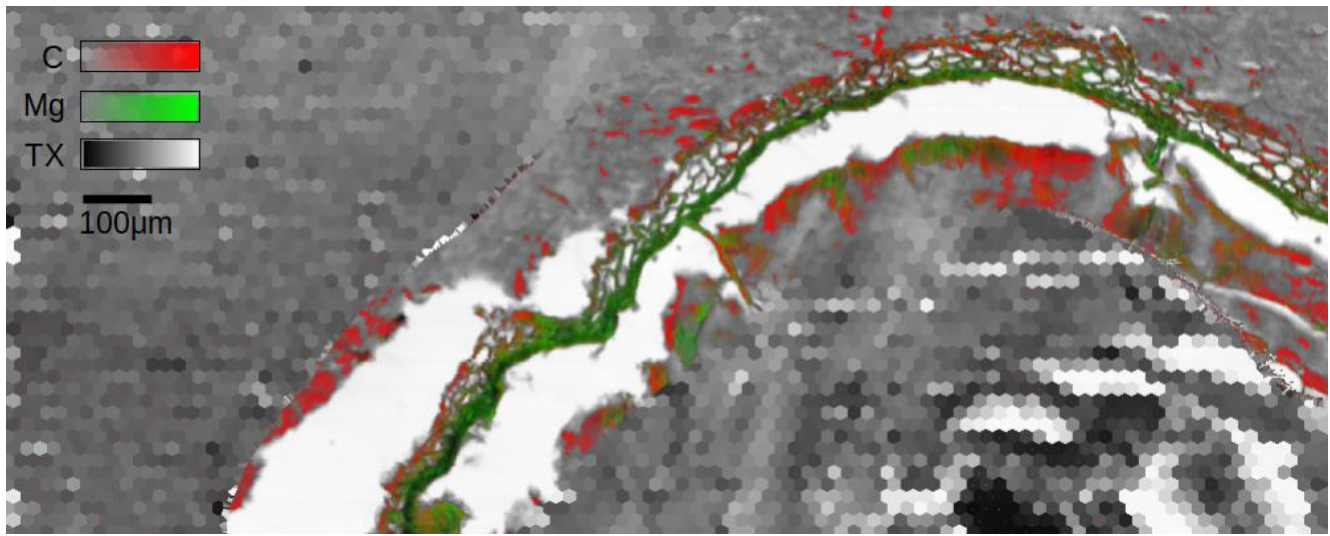

**Figure 5.** 0.4MPixel Sparse scan and Masking combined with a Conditional scan and multimodal acquisition on a cane root section sample. The STXM data (grey) (2000μm x 800μm) are acquired with a dense sampling (2μm step size) along the border of the cane root section while in a sparse way (20μm step size) in the remaining masked areas. The XRF data have been collected in a similar reduced manner (C in red, Mg in green) and are displayed overlapped with the STXM image. A traditional/complete acquisition (2μm step size) on the total area would require a measurement time 3 times longer.

**Conclusions**

This paper suggests a series of methods for performing XRF scans in dynamic ways. It is a Compressive Sensing approach where the data are collected in a reduced manner and later are reconstructed to standard representations. This paves the way for larger and faster scans enabling new science. Through a series of beamtime experiments and suitable software and instrumentation modifications, the feasibility of such an approach was demonstrated. The proposed techniques are presented in three main categories; Sparse scans & Masking, Conditional scans & multimodal acquisitions, and Dynamic XRF scanning. Examples of reduced raw data and final reconstructions are shown and FAIR data and codes are set accessible to the community through a public repository [doi:10.5281/zenodo.3688316]. Finally certain future perspectives have been presented and hint on the use of Machine Learning for the decision-making and reconstruction of missing data component of the methods. The proposed techniques can be combined with other advances on instrumentation and detectors but it should be noted that they introduce an additional challenge on the post-processing reconstruction phase of the reduced data. The expected impact of this research is the use of powerful synchrotron LEXRF for the study of a high number of biological samples of large areas at very high resolutions. Naturally the potential applications can be across different fields from biology to material science in different energy ranges and potentially could include other scanning microscopy techniques (SEM, FTIR, Ptychography etc).

**Materials and Methods**

**Experimental**
The LEXRF experiments were performed at the TwinMic beamline[12]. The end-station was operated in Scanning Transmission Mode (STXM) where the monochromatic beam is focused on the sample plane by means of zone plate diffractive optics while the specimen is raster-scanned across the microprobe[18]. For the present experiments a gold 600 μm diameter zone plate with 50nm outermost zone was used producing a microprobe with 1 μm diameter size at 2 keV. The samples were raster-scanned with a step size of 20 or 2 micron.
The specimens shown in Figures 1 and 2 are 10 micron thick sections of foraminifera shells deposited on ultralene foils. The sample depicted in Figure 5 is a 10 micron thick cross section of a cane root deposited on a ultralene foil.

**Software**
In order to develop, test and deploy different scanning strategies, major changes in the existing beamline end-station control software were required. Typical software for synchrotron XRM end-stations allows only for raster rectangular acquisitions with few exceptions on Ptychography setups, but even those have predefined sets of scan positions and acquisitions parameters. A novel system was developed by engineers among the authors of this paper. It permits for non-regular positions and dynamic parameters that can all be changed during the scan at per-pixel basis. This system includes an advanced workflow manager (DonkiOrchestra[13]) which allows for easy and flexible changes of the data acquisition process. Other institutes are using similar systems (see Bluesky [https://nsls-ii.github.io/bluesky/ accessed: 25/02/2020]) thus the implementation of the proposed techniques also in other facilities should be quite easy. The data were stored in HDF5 files of non-standard structure and were later reconstructed in traditional dense data structures that can be stored in common formats (i.e. NeXus) and processed with popular analysis software (e.g. PyMca[17]). Both the acquisition framework and the reconstruction software were done in Python. The experimental data and their reconstructions are available at a public repository [doi:10.5281/zenodo.3688316] respecting the FAIR principles.

**Supplementary Information**

**Compressive Sensing for Dynamic XRF Scanning**

George Kourousias, Fulvio Billè, Roberto Borghes, Antonio Alborini, Simone Sala, Roberto Alberti and Alessandra Gianoncelli

(1)

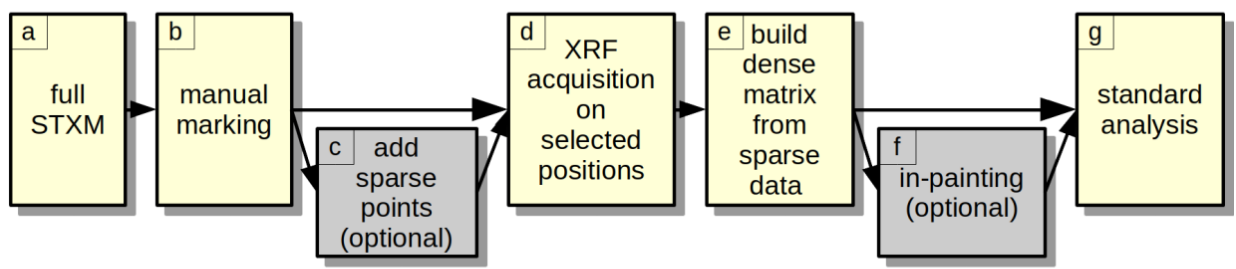

(2)

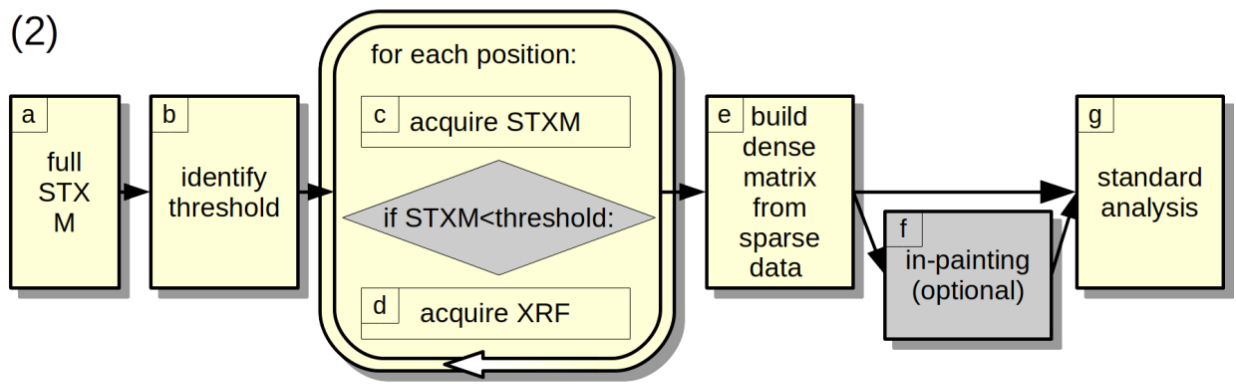

(3)

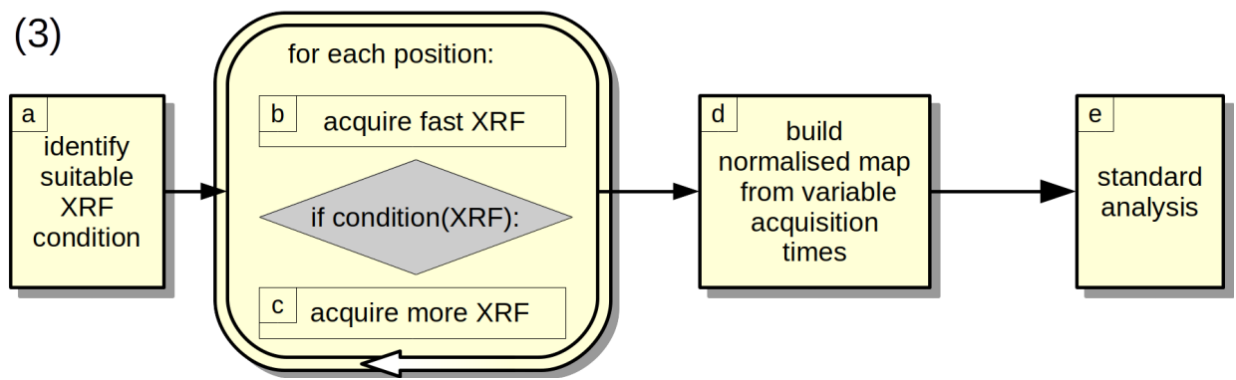

**S1.** *Schematic workflows for (1) Sparse scans and Masking, (2) Conditional and Multimodal acquisitions, and (3) Dynamic scans.*

A procedural description of the methods presented in this manuscript may provide an overview of the workflow and assist their implementation in other beamlines. The *sparse scans and masking* method (Results and Methods, section 1) requires a full STXM scan (s1.1a) which is used for the spatial selection of a suitable sub-ROI. This manual selection may be irregular and in TwinMic is done with a Wacom graphics tablet. Optionally, a suitable sparse pattern can be added outside the ROI in order to assist the in-painting/filling-the-missing-points method (s1.1c). The ROI, possibly expanded by the sparse pattern, results in a reduced set of scan points used by the acquisition system leading to a sparse set of XRF data (s1.1d). These data through computation are suitably reconstructed in a regular matrix (s1.1e) of the scanned area but including the missing values (Fig1e). Optionally, a suitable in-painting method (s1.1f) is used to reconstruct those missing values. The data at this stage can be further processed with any typical XRF workflow (s1.1g) thus calibrated, filtered, fitted, quantised etc.

The following workflow (s1.2) is for *conditional scans and multimodal acquisitions* (Results and Methods, section 2) where a secondary imaging technique is the decisive factor for XRF acquisition. Like for the previous workflow, it starts with an STXM acquisition (s1.2a) but its purpose is not to identify valid positions but only an x-ray transmission threshold (s1.2b). The threshold should discriminate the sample from its support or identify the most significant absorbing areas. The threshold can be derived from a representative area of the sample and then applied to a much larger scan. Once the threshold has been selected, for each position, the transmission signal is acquired (s1.2c) and according to a conditional comparison with the previously established threshold, a longer XRF acquisition is triggered or not (s1.2c). This results in a sparse set of data that are reconstructed to a matrix which may be in-painted and subsequently analysed with standard methods like in the previous workflow (s1.2e,f,g).

The final and most challenging workflow of dynamic scans (s1.3) (Results and Methods, section 3), does not require an additional technique like STXM (s1.1a, s1.2a) but assumes a suitable XRF condition (s1.3a) like a simple XRF signal threshold for a specific chemical element (as used in the presented results in Figure 3) or an intelligent ratio between different peaks etc. Following this, for each position in the scanned area, a fast XRF acquisition (s1.3b) is used together with the previously mentioned criteria (i.e. Na XRF levels) to decide dynamically whether to acquire a longer XRF signal (s1.3c) on the same position for better statistics. This results in a set of XRF data collected at variable exposure times that after a suitable normalisation can be reconstructed to a regular matrix (s1.3d).
Note that these three generic workflows can be easily combined resulting in an adaptable acquisition scheme.

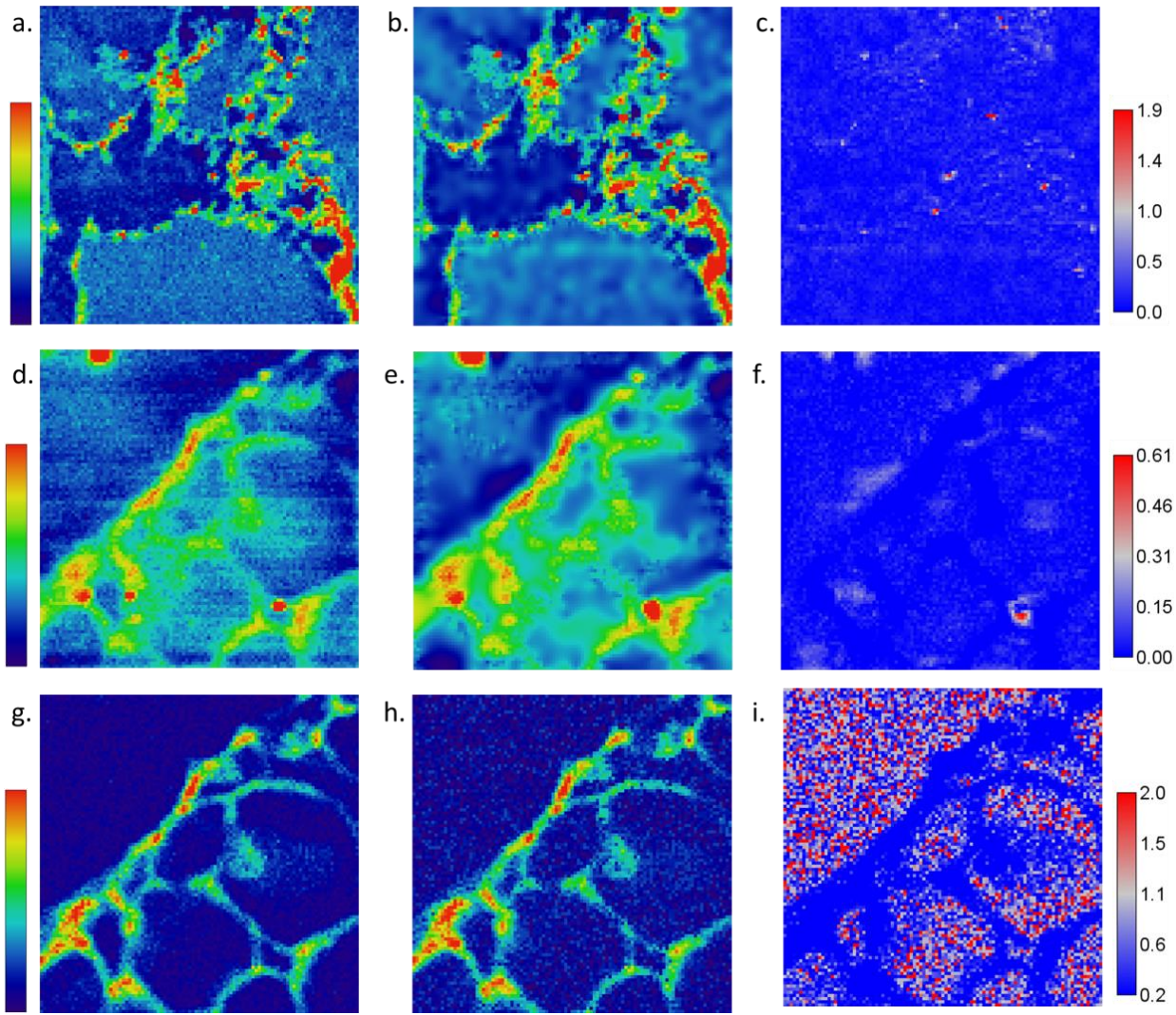

**S2.** *Reconstructed image evaluation of error. Relative error (c,f,i) between full XRF maps (a,d,g) and their corresponding reconstructions (b,e) from sparsely acquired data or dynamically varying acquisition (h). Panels (a,b,c) supplement Figure 1 on Sparse Scans and Masking, while (d,e,f) supplement Figure 2 on Conditional and Multimodal Acquisitions. Panels (g,h,i) refer to Figure 3 on Dynamic Scans where (g) is a full constant time acquisition and (h) is a variable one with better statistics in the regions of interest. The Error (c,f,i) is calculated as* `|recostructed-original|/original`.

The suggested methods as presented in their workflows consist of steps like those of mask definition, threshold identification and choice of signal logic (s1.1b,s1.2b,s1.3a) in combination with the reconstruction methods of XRF maps from sparse data (s1.1-2e-f, s1.3d-e). These steps can impact the quality of the reconstructed XRF map. On top of that, factors like mechanical instabilities (motor imprecisions and thermacal drifts) and x-ray beam fluctuations (position and intensity) add to the problem. Note that the final quality of this lossy process needs to be as high as possible aiming at reducing the losses at specific regions of interest while acquiring reduced data. Other than the encouraging empirical evaluation by an expert, a basic error measurement technique was proposed during the refereeing of this manuscript. The error is calculated as `|reconstructed-original|/original`.

Other evaluation approaches have been implemented as well, including operating in logarithmic scales, pixel-wise standard deviation and structural similarity (SSIM), all of them resulting in comparable results. In specific, for data shown in Figure 1 representing the method of Sparse Scans and Masking the in-painted reconstruction (s2b) was compared to a typical full scan (s2a) resulting in an error map (s2c). Since these data, full and sparse (s2a,b) are collected subsequently and the mask is based on an initial full STXM acquisition (s1a), any sample stage instabilities create displacements that can be noticed after a pixel-wise comparison in the red values (>1) in their error map (s2c). Similarly, for the case of Conditional and Multimodal Acquisitions, for the maps of Figure 2 it was compared the in-painted reconstructed map (s2b) to the full one (s2a). While this technique is not requiring the *a priori* definition of a mask, it does require the identification of a transmission threshold (s1.2b). This makes it robust to mechanical instabilities since even if the sample is moved, the process decides point-by-point whether to acquire or not (s1.2c,d). This is also reflected on the error map (s2f) that is substantially lower than that of the Sparse scanning and Masking method (s2c). With a mean error of 0.026 and the qualitative evaluation of experience microscopy operator, it demonstrates a good case of compressive sensing.

An interesting remark from these experimental data is the error visible on the lower right section of the error map (in red) that results in a higher XRF signal on the reconstructed map (s2e) compared to the original one (s2d). After careful observation, it appears that this region has been mostly excluded from the STXM threshold (Fig2b,c,d) even though it has a strong XRF emission. This suggests that STXM may not always reveal the XRF regions of interest but also that the in-painting methods may overcompensate and introduce additional signal/artefacts through wrong interpolation.

**Acknowledgements**
The authors would like to thank Maya Kiskinova from Elettra Sincrotrone Trieste (Italy) for insights on further use of the methods in pump-and-probe experiments, Alfonso Franciosi from Elettra Sincrotrone Trieste (Italy) for raising interest on large scale imaging and Francesco Guzzi from Elettra Sincrotrone Trieste and University of Trieste (Italy) for investigating Machine Learning in-painting techniques. The authors are grateful to Giovanni De Giudici and Daniela Medas from University of Cagliari (Italy) for providing the foraminifera and cane root samples, and Peter Kopittke from Queensland University for the soybean root samples. TwinMic (Elettra) 20180178 and 20190372 beamtimes are acknowledged.

**Competing interests**
The authors declare no competing interests.

**Contributions**
G.K. conceived the experiments and methods, F.B. and G.K. designed the data analysis software, R.B. developed the acquisition software, A.A. and R.A. provided the detector processing electronics, A.G. supervised the experiments and with S.S. provided support in the beamline. All authors participated in the experiments and contributed to the manuscript.